\documentclass[preprint,preprintnumbers,amsmath,amssymb]{revtex4}

\usepackage{graphicx}
\usepackage{dcolumn}
\usepackage{bm}
\usepackage{amsmath, amsthm, amssymb}
    
\usepackage{tabularx}
    \newcolumntype{Y}{>{\centering\arraybackslash}X}

\usepackage{delarray}

\begin{document}

\preprint{}

\title{Electrical Characterization of PbZr$_{0.4}$Ti$_{0.6}$O$_3$ Capacitors}

\author{P. Zubko}
  \email{pz212@cam.ac.uk}
\author{D. J. Jung}%
  \altaffiliation[Also at ]{Advanced Technology Team, Semiconductor R\&D Center, Samsung Electronics Co. LTD,  San\#24, Nongseo-Dong, Giheung-Gu, Yongin-City, Gyunggi-Do, South Korea}
\author{J. F. Scott}
\affiliation{%
Centre of Ferroics, Department of Earth Sciences, University of Cambridge, Cambridge CB2 3EQ, United Kingdom 
}%

\date{\today}

\begin{abstract}
We have conducted a careful study of current-voltage (\emph{I-V}) characteristics in  fully integrated commercial PbZr$_{0.4}$Ti$_{0.6}$O$_3$  thin film capacitors with Pt bottom and Ir/IrO$_2$ top electrodes. Highly reproducible steady state \emph{I-V} were obtained at various temperatures over two decades in voltage from current-time  data and analyzed in terms of several common transport models including space charge limited conduction, Schottky thermionic emission under full and partial depletion and Poole-Frenkel conduction, showing that the later is the most plausible leakage mechanism in these high quality films. In addition, ferroelectric hysteresis loops and capacitance-voltage data were obtained over a large range of temperatures and discussed in terms of a modified Landau-Ginzburg-Devonshire theory accounting for space charge effects. 
\end{abstract}

\pacs{Valid PACS appear here}
\maketitle

\section{Introduction}
It has been known for a long time that the description of ferroelectric oxides as insulators becomes less and less valid as their dimensions are reduced. The conductivity of thin films is often modelled using well established semiconductor theories and  concepts for new devices such as resistive memories \cite{Contreras, MeyerKolstedt} are already emerging which exploit the semiconducting nature of thin ferroelectrics. The frequently observed deformation of hysteresis loops by leakage currents, which masks the true ferroelectric properties of the system, apart from being a nuissance to researchers,  is not directly a problem for non-volatile memory applications. However, charge injection associated with finite conductivity is believed to be involved in accelerating fatigue as well as raising other reliability issues. The increased power consumption and heating due to leakage are also undesirable. Leakage measurements are therefore a crucial part of any electrical characterization of a ferroelectric device. These are often performed using automated systems designed for semiconductor characterization. However, as shown in this article, great care should be taken if one wants to obtain true steady state current-voltage (\emph{I-V}) data due to the the much slower dielectric relaxation in ferroelectrics compared with usual semiconductors as well as resistance degradation at high fields. Such artefacts in \emph{I-V} measurements can lead to incorrect interpretation of the conduction mechanism.

This paper reports a detailed study of the \emph{I-V} characteristics of Samsung's lead zirconium titanate (PZT) capacitors. Together with capacitance-voltage (\emph{C-V}) measurements  and ferroelectric hystesis (\emph{P-V}) loops, the data are analyzed in the light of various conduction models with the aim of understanding the leakage mechanism.

\section{Experimental}
The 144nm PbZr$_{0.4}$Ti$_{0.6}$O$_3$ films were deposited by a sol-gel method at Samsung, using a high purity (99.9999\%) precursor, on a [111] oriented Pt bottom electrode, resulting, after annealing, in an almost epitaxial interface between the metal and the columnar PZT grains. For each capacitor the top Ir/IrO$_2$ electrode was patterned as a two-dimensional array of interconnected $\sim$13$\times 13\mu$m$^2$ squares with a total area  $A=5.31\times 10^{-5}$cm$^2$. The capacitor structures were fully integrated and embedded in Si, with  various metallic layers connecting the top and bottom electrodes to the external tungsten contact pads. Current-voltage-time (\emph{I-V-t}) measurements were performed using the Agilent 4155C semiconductor parameter analyser, dynamic capacitance measurements were obtained using the HP4192A impedance analyzer and the Radiant Technologies Precision Pro tester was used for ferroelectric hysteresis loops.

\section{Results}
\subsection{Current-voltage characteristics}
\begin{figure}[t]
\centering
\scalebox{0.75}{\includegraphics{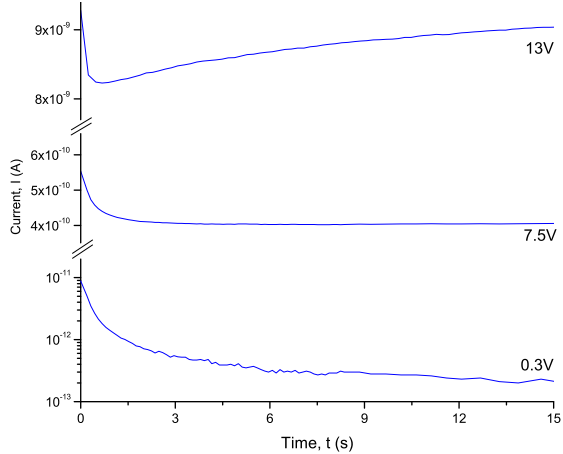}}
\caption{Typical current transients obtained at $42^{\circ}$C at various biases. Note the change of scale from logarithmic  at low field to linear at higher biases.}
\label{Transients:OK}
\end{figure}
The \emph{I-t} response at $42^{\circ}$C of our capacitors upon a stepwise application of a constant bias is shown in figure~\ref{Transients:OK} and is typical of leakage behaviour of ferroelectric thin films. At short times the behaviour is dominated by dielectric relaxation which often follows the empirical Curie-von Schweidler law ($I\propto t^{-n}$ with $0<n<1$). This is then followed by a region of steady state leakage (in this case most clearly visible  for the intermediate bias), which may or may not be observed depending on the point at which the current starts to increase again due to time-dependent dielectric breakdown or resistance degradation. It is common practice when measuring \emph{I-V} characteristics to choose a waiting time (between the application of the bias and the current measurement) which is longer than the dielectric relaxation time at intermediate biases. However, as is clear from figure~\ref{Transients:OK}, this can lead to an overestimate of the current at high biases due to resistance degradation and at low biases due to dielectric relaxation, sometimes leading to apparent negative differential resistivity or ohmic-like behaviour. 

To obtain true steady state current readings and avoid unnecessary stress to the sample, we performed \emph{I-t} measurements at each voltage until a clear plateau in the current was reached. In a few cases at high biases and high temperatures where no clear plateau could be resolved before the onset of resistance degradation, the minimum value of the current was used. The tungsten pads were contacted using 10$\mu$m diameter tungsten probe tips, which were found to be more effective in achieving a good contact and penetrating through the thin tungsten oxide surface layer than thinner tips. Due to the high homogeneity of the films and the almost perfect reproducibility of all electrical measurements, the quality of contact could be checked by looking at the hysteresis loops, and in particular at the coercive voltages, which were found to be very sensitive to any contamination of the tungsten pad surfaces. Prior to taking any measurements the samples were poled at three times the coercive field (for $\sim$60s) and at the highest measurement bias (for 10-20s, depending on the temperature) to avoid any diplacement currents and internal field changes due to the reversal of ferroelectric polarization when measuring the \emph{I-V} characteristics. In this way the forward and reverse bias \emph{I-V} measurements were obtained for a range of temperatures; we will define forward bias as the case when the top IrO$_2$ electrode is at a higher potential than the bottom Pt electrode.
\begin{figure}[h]
\centering
\scalebox{0.8}{\includegraphics{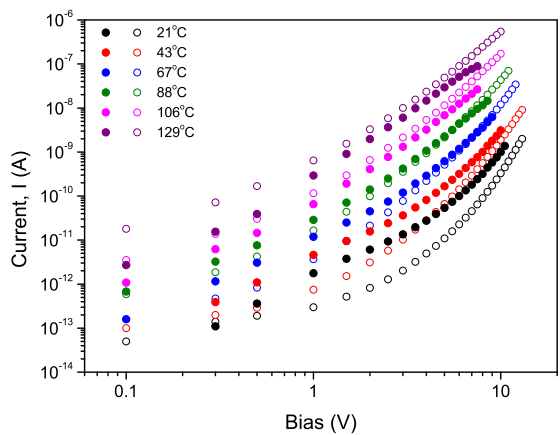}}
\caption{Forward (open circles) and reverse (filled circles) bias \emph{I-V} at different temperatures.}
\label{FwdAndRvsIV:OK}
\end{figure}

Figure~\ref{FwdAndRvsIV:OK} shows that at intermediate temperatures the \emph{I-V} are quite symmetric apart from the very low field region below $\sim 1$V. At low and high temperatures, however, some asymmetry is seen. Although different capacitors were used to obtain the forward  and reverse bias data, this cannot account for the observed asymmetry due to the high reproducibility of the measurements  over many samples, especially in forward bias. The asymmetry is quite low, well below an order of magnitude for almost all biases. It is also interesting to note that whereas in low temperature measurements the reverse bias currents are higher, they are lower at high temperatures; this is not expected if the asymmetry is due to differences in the interfaces, such as different barrier heights, at the top and bottom electrodes.  In general, reverse bias currents were  somewhat less reproducible and less stable, with breakdown occuring at lower biases. For this reason we focused more on forward bias data for the quantitative analyses, althogh some reverse bias \emph{I-V} will also be discussed. 

In the hope of  understanding the mechanism responsible for conduction in these films we have looked at a number of models commonly used to describe carrier transport in ferroelectrics and tried to fit their predictions as quantitatively as possible to our own data. In the following subsections we therefore present the same set of results analyzed in terms of these different models.

\subsection*{Space charge limited (SCL) conduction}
When the internal field is dominated by space charge (either from free or trapped carriers)  the current density $J$ is, in general, expected to have a power-law dependence on voltage, $J\propto V^n$. In the case of discrete traps a quadratic voltage and cubic thickness ($d$) dependence is expected \cite{HwangKao}
\begin{equation}
J=\frac{9}{8}\theta_f\epsilon\epsilon_0\mu \frac{V^2}{d^3}\nonumber,
\end{equation}
where $\mu$ is the carrier mobility, $\epsilon$ the relative dielectric constant and $\theta_f$ the fraction of injected carriers that are free (i.e. not trapped); in the trap-free limit $\theta_f=1$. Unfortunately, since these are commercial Samsung films of a fixed thickness, we have to rely purely on the voltage dependence when analyzing our data.

From figure~\ref{FwdAndRvsIV:OK} it is possible to identify, for reverse bias data in particular, regions which look linear on the log-log plot and could therefore satisfy the power-law dependence. At low negative biases the room temperature exponent $n$ was found to be 2.1, which is compatible with the trap-free or discrete trap case.  However, an unexpected temperature dependence was found, with $n$ decreasing steadily to 1.65 at the highest temperature. The high-field exponents were 	in the range 3.0--4.8 for reverse bias and 3.5--6.9 for forward bias data respectively. Such high exponents are not common to SCL currents, and are certainly not possible for discrete traps. Values of $n>2$ are possible if traps are distributed within the bandgap and they are expected to be temperature dependent \cite{HwangKao}. In the case of an exponential trap distribution characterised by a temperature $T_1$
\[n-1=\frac{T_1}{T}\]
and for a Gaussian trap distibution with a standard deviation $\sigma_t$  we expect 
\[(n-1)^2=1+\frac{2\pi\sigma_t^2}{16k^2T^2}.\]
However, neither of these relationships were well satisfied by our data. We therefore believe that SCL conduction is not likely to be the dominant leakage mechanism in our films, though other authors \cite{ Shin_SCLC, Cho_Jeon_SCLC} have tried to explain their data using this model.

\subsection*{Schottky thermionic emission}
The slight asymmetry between forward and reverse bias \emph{I-V} suggests that the interface properties might play a role and we therefore consider Schottky thermionic emission as a possible conduction mechanism. In this case the sample is considered as two back-to-back Schottky diodes, with the current controlled by the  field-induced lowering of the orginal zero-field potential barrier height $\phi_{B0}$ at the reverse (or forward) biased electrode  when electrons (or holes) are the majority carriers. The current density is given by \cite{Sze}
\begin{equation}
J=A^*T^2\exp(-q\phi_{B0}/kT)\exp(\beta\sqrt{E_m})
\label{Schottky:OK}
\end{equation}
with the Richardson constant $A^*=4\pi em^*k^2/h^3$ related to the effective carrier mass $m^*$, and $\beta=(e/kT)(e/4\pi\epsilon\epsilon_0)^{1/2}$; $k$ and $h$ are the Boltzmann and Planck constants  and $q$ and $e$ are the carrier and electron charges respectively. The important parameter in equation~\ref{Schottky:OK} is the field $E_m$ at the potential maximum which determines the barrier lowering. We consider two cases.
\subsubsection*{Case 1: uniform field (full depletion)}
Bulk ferroelectrics are good insulators and  even in thin films it is common to assume that $E=V/d$. If the films are semiconducting this corresponds to the case of full depletion, which is expected for low concentrations of ionised dopants.
\begin{figure}[h]
\centering
\scalebox{0.8}{\includegraphics{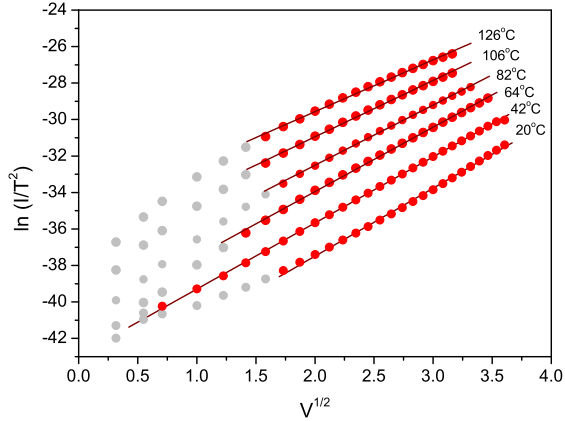}}
\caption{Forward bias data represented on the standard Schottky plot. The data which deviate from the linear fit are shown in light grey.}
\label{SchottkyFullDepl:OK}
\end{figure}
Figure~\ref{SchottkyFullDepl:OK} shows that, especially at higher fields, the data satisfy the Schottky equation quite well; but we also have to see if the physical parameters that can be extracted from the data are sensible. The relative dielectric constant  was calculated from slopes of the linear fits in figure~\ref{SchottkyFullDepl:OK} to be between 0.9 and 1.1 over the temperature range shown. The physical range for $\epsilon$ is limited from one side by the optical dielectric constant $\epsilon_\infty\approx 6$ \cite{Eps_Infty} and from the other side by the static dielectric constant $\epsilon_{dc}$, expected to be around 400 \cite{Haun} for this composition along [111]. The calculated value of around $\epsilon\sim 1$ is clearly outside this range. From the temperature dependence of the intercepts the  Richardson constant and effective carrier mass were found to be 2.4~Acm$^{-2}$K$^{-1}$ and 0.02m$_e$ respectively, where m$_e$ is the free electron mass. These values also seem somewhat low compared to those previously reported for perovskites \cite{EffectiveMass}. A barrier height of 0.9eV was also calculated, which is plausible from band diagram considerations.

\subsubsection*{Case 2: partial depletion}
In more semiconducting films, the dopant concentration $N_D$ may be sufficiently high to allow the applied voltage to be dropped across a thin depletion layer near the reverse   biased electrode (for an $n$-type or forward biased for a $p$-type material), resulting in a linear field profile in the sample and a modified value for $E_m$ which also depends on the built-in bias $V_{bi}$ \cite{Sze}
\begin{equation}
E_m=\sqrt{\frac{2qN_D}{\epsilon_{dc}\epsilon_0}\left(V+V_{bi}-\frac{kT}{q}\right)}.
\label{FieldPartDepl:OK}
\end{equation}

The $kT/q$ term originating from the free carrier contribution to the space charge within the depletion layer may usually be neglected.

Recently Pintilie \emph{et al.} \cite{Pintilie} have analysed their epitaxial single crystal PZT data using the partially depleted thermionic emission model,  modified to account for the ferroelectric polarization, and have shown that they fit the $\log (J/T^2)\sim (V+V_{bi})^{1/4}$ law. We performed a similar analysis on our films, fitting the \emph{I-V} to the quarter-power law assuming various values of $V_{bi}$ and extracting  the physical parameters from the slopes and intercepts of our fits.

\begin{figure}[h]
\centering
\scalebox{0.8}{\includegraphics{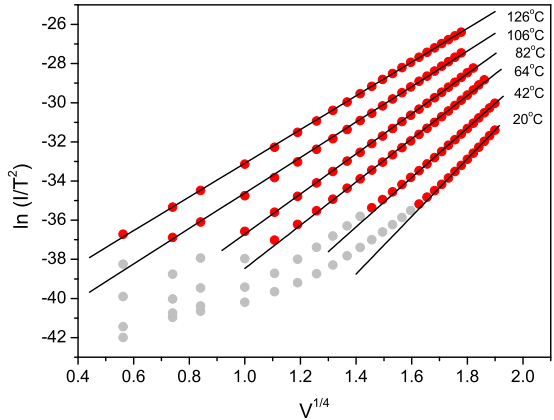}}
\caption{Forward bias \emph{I-V} fitted to Schottky thermionic emission model under partial depletion with zero $V_{bi}$.}
\label{PartDeplZeroVbi:OK}
\end{figure}

We found that the quality of the fit could be adjusted by varying $V_{bi}$, with the high temperature data fitting best for low or negative values of $V_{bi}$ and the low temperature data  giving the best fits assuming a high positive built-in bias. From the temperature dependence of the intercepts of $\log (J/T^2)$ vs. $(V+V_{bi})^{1/4}$ plots, the effective carrier masses and barrier heights were extracted and are shown in table~\ref{Parameters:OK} for different assumed values of $V_{bi}$.
\begin{table}[h] 
\begin{tabularx}{\linewidth}{|
>{\setlength{\hsize}{1.0\hsize}}Y
>{\setlength{\hsize}{1.0\hsize}}Y
>{\setlength{\hsize}{1.0\hsize}}Y|}
\hline
$V_{bi}$ (V) & $m^*/m_e$ & $\phi_{B0}$ (eV) \\
\hline

-0.05	&	20		&	1.34	\\
0		&	2.1		&	1.28	\\
0.1		&	0.10	&	1.21	\\
0.2		&	0.010	&	1.16	\\
0.4		&	0.00026	&	1.09	\\
0.6		&	1.27$\times 10^{-5}$	&	1.02	\\
\hline

\end{tabularx} 
\caption{Effective carrier masses are highly dependent on the $V_{bi}$ value assumed for the fit. It should also be noted that, due to the exponential dependence of $m^*$ values on the intercept,  they are very sentitive to small errors.} \label{Parameters:OK}
\end{table}
The effective carrier masses in perovskites are expected to be of order a few $m_e$ so the most physically plausible value for $V_{bi}$ is close to  zero. The $\log J$ vs. $(V+V_{bi})^{1/4}$ plot with $V_{bi}=0$ is shown in figure~\ref{PartDeplZeroVbi:OK} and, it is clear that only the high temperature data gives a good fit in this case.

Provided the  mobility of the carriers is not too low, so that they only spend a short time near the potential barrier maximum, $\epsilon$ in equation~(\ref{Schottky:OK}) should be replaced by $\epsilon_{\infty}$. Substituting expression~(\ref{FieldPartDepl:OK}) for $E_m$ into (\ref{Schottky:OK}) shows that the slopes in figure~\ref{PartDeplZeroVbi:OK} allow the product $N_D/\epsilon_\infty ^2\epsilon_{dc}$ to be calculated. Assuming $\epsilon_\infty=6$ and $\epsilon_{dc}\approx 400$  gives a rough estimate for the concentration of ionised impurities $N_D\approx 10^{20}$--$10^{22}$cm$^{-3}$.

\subsection*{Poole-Frenkel conduction}
\begin{figure}[h]
\centering
\scalebox{0.8}{\includegraphics{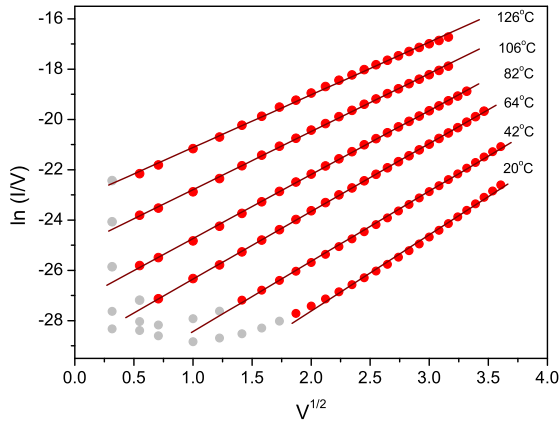}}
\caption{Forward bias data showing a very good fit to the Poole-Frenkel model  over two decades in bias.}
\label{PooleFrenkelFig:OK}
\end{figure}

In insulating samples, field-enhanced emission of carriers from traps can lead to Poole-Frenkel currents, which have a functional field dependence similar to Schottky emission currents with
\begin{equation}
J=\sigma_0E\exp\left(-\frac{E_t}{kT}\right)\exp(\beta_{PF}\sqrt{E}),
\label{PooleFrenkel:OK}
\end{equation}
where $E_t$ is the trap energy, $\beta_{PF}=2\beta$ and $\sigma_0$ is the sample dependent zero-field conductivity. Poole-Frenkel conduction is a bulk mechanism and therefore it is usual to assume $E=V/d$. Thickness dependence can be very useful in discriminating between bulk and interface effects, however, as already mentioned such a study was not possible in our case. Thus to test the Poole-Frenkel model we have to rely purely on the fitting of the \emph{I-V} curves. Figure~\ref{PooleFrenkelFig:OK} shows that the data fit equation~(\ref{PooleFrenkel:OK}) extremely well over the whole two decades in voltage at almost all temperatures. The relative dielectric constant extracted from the slopes was found to be 6.3 at room temperature and  in the range of 6.3--7.4 for the six curves shown with no specific temperature dependence. These values are very close to the optical dielectric constant. The trap energy was found to be 1.0eV.
\subsection{Capacitance-voltage}
\begin{figure}[h]
\centering
\scalebox{0.8}{\includegraphics{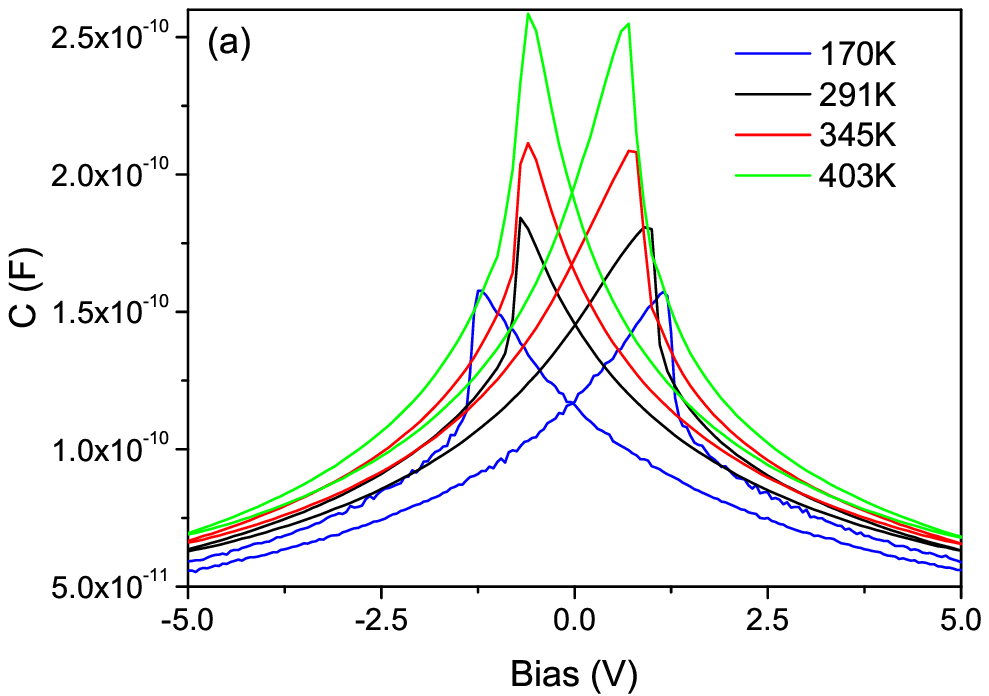}
\includegraphics{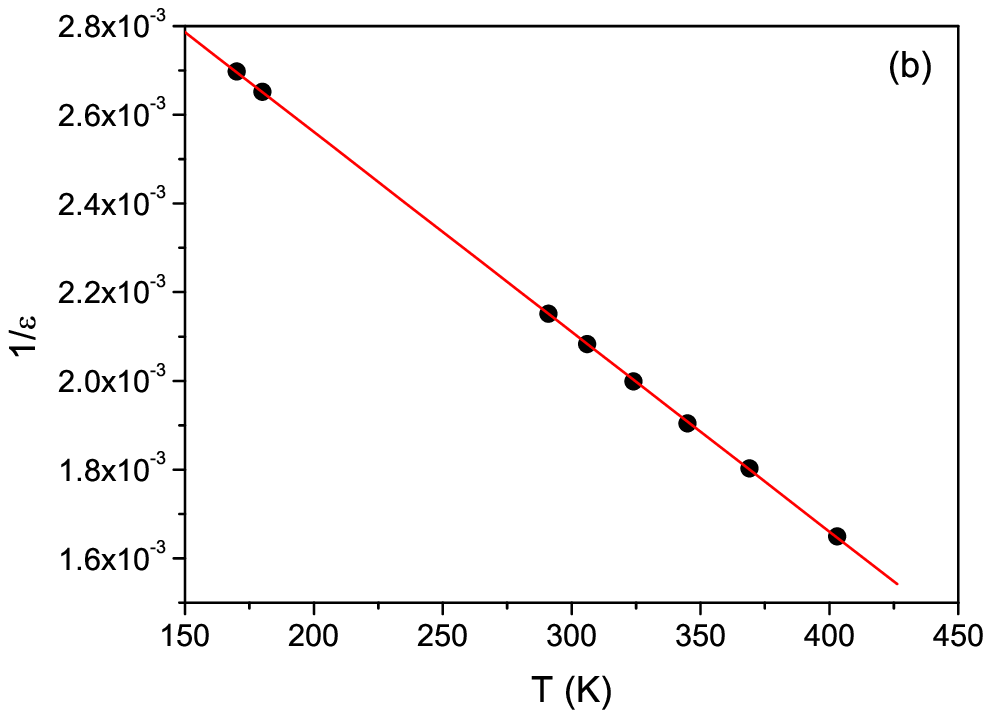}}
\caption{(a) Capacitance-voltage curves at different temperatures. (b) Curie-Weiss plot for the inverse of the relative dielectric constant.}
\label{C_V:OK}
\end{figure}

The capacitance was measured as a function of dc bias using a 100kHz 25mV tickle voltage at high and low temperatures with typical curves shown in figure~\ref{C_V:OK}(a). The room temperature zero-bias dielectric constant extracted from the capacitance measurement via $\epsilon=Cd/A\epsilon_0$ was found to be  around 460. The a-axis and c-axis dielectric constants for bulk PbZr$_{0.4}$Ti$_{0.6}$O$_3$ are 498 and 197 respectively \cite{Haun}. We believe our samples to be preferentially [111] oriented and therefore the extracted dielectric constant is not far from the value of 398 expected for perfectly  [111] oriented bulk PZT. Figure~\ref{C_V:OK}(b) shows a Curie-Weiss plot for the inverse zero-bias dielectric constant obtained from the capacitance measurements. From the gradient of the plot and equation~(\ref{InvEps:OK}) a Curie constant $C=4.44\times 10^5$K$^{-1}$ was calculated.

If the samples were partially depleted,  a change in bias would result in a change of  the depletion widths and consequently of the depletion layer capacitance. The concentration profile of ionised impurities   can therefore in principle be calculated from the \emph{C-V} measurements using
\begin{equation}
N_D(V)=\frac{2}{q\epsilon_{dc}\epsilon_0\frac{d}{dV}\left(\frac{1}{C^2}\right)},
\end{equation}
where each value of $V$ probes $N_D$ at a particular depth within the film. In practice, however, only the mobile carriers respond to the ac tickle voltage used for the \emph{C-V}  measurements, thus only mobile carrier concentrations  ($n$ or $p$) can be obtained. Using this method and again assuming $\epsilon_{dc}=400$ we obtained a value of $\sim 3\times 10^{18}$cm$^{-3}$ for the mobile carrier concentration. 

\subsection{Ferroelectric properties}
\begin{figure}[h]
\centering
\scalebox{0.8}{\includegraphics{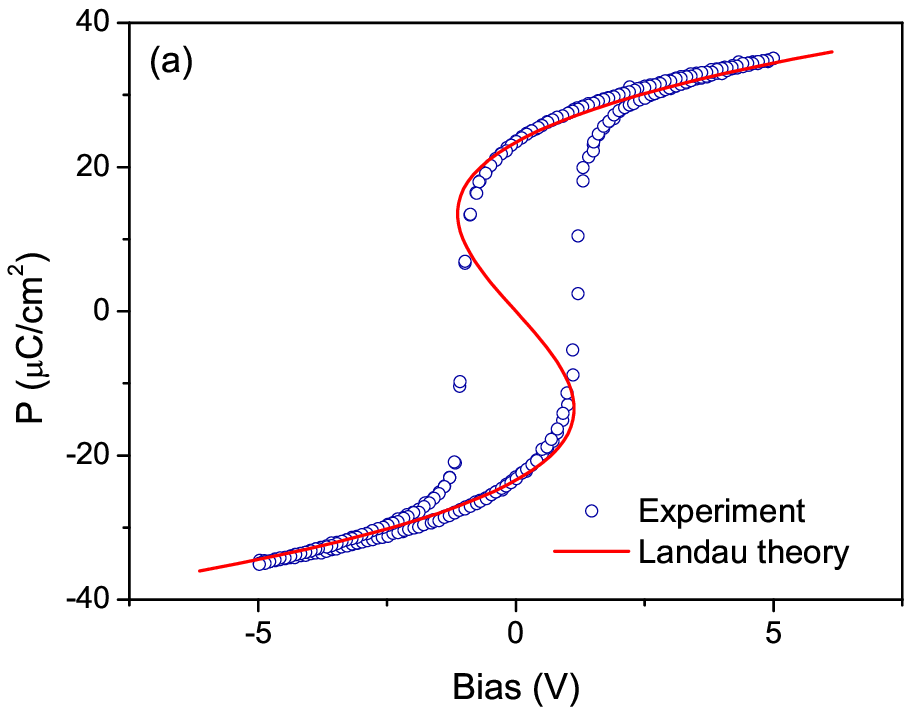}
\includegraphics{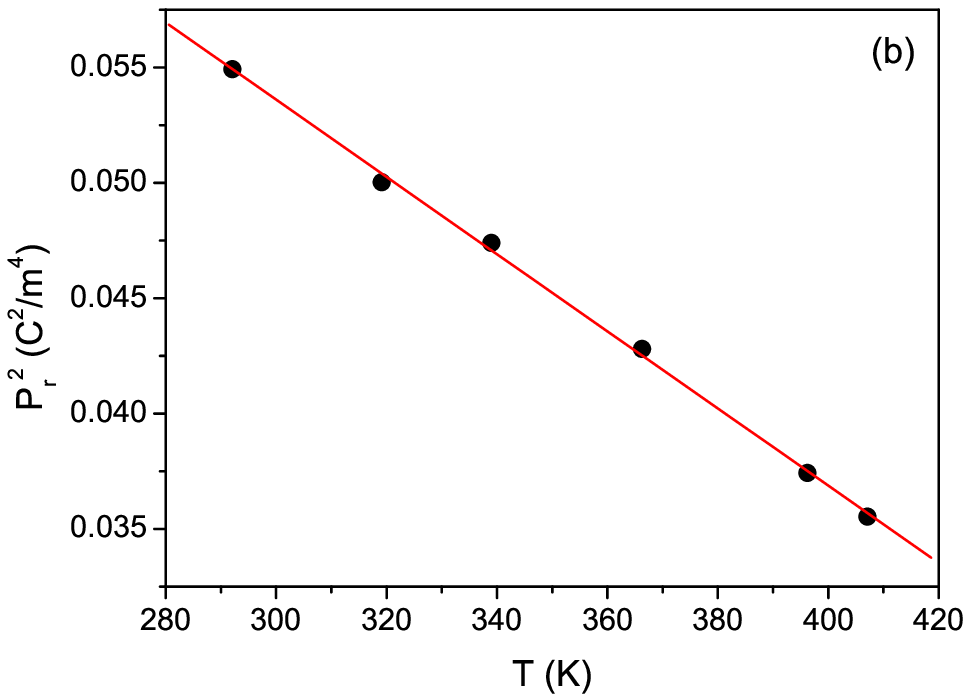}}
\caption{(a) Room temperature \emph{P-V} loop; (b) Square of the remanent polarization $P_r$ as a function of temperature.}
\label{P_V:OK}
\end{figure}
A typical ferroelectric hysteresis loop is shown in figure~\ref{P_V:OK}a, illustrating the very well defined coercive voltages  and  a symmetric shape with very little imprint. The remanent polarization of 23$\mu$C/cm$^{2}$  is significantly smaller than the 33$\mu$C/cm$^{2}$ expected for bulk [111] PZT of the same composition. 
The capacitors are known to be under compressive strain, so strain effects cannot be responsible for the lowering of the polarization as they would increase it. It was observed that $P_r$ progressively decreases after each etching step during the fabrication process and must therefore be associated with introduction of defects such as space charge or perhaps passive interfacial layers.

Within the framework of Landau-Ginzburg-Devonshire (LGD) theory, for ferroelectrics with a second order phase transition and with polarization low enough that the $P^6$ terms in the free energy expression can be neglected, the hysteresis loops are given by a cubic equation
\begin{equation}
E_3=\alpha P_3 + \beta P_3^3,
\label{cubic:OK}
\end{equation}
where the field  and polarization are assumed to be along the c-axis (hence the subscript 3)and $\alpha$ and $\beta$ are the standard bulk LGD coefficients. Provided the coefficient coupling the polarizations along the different axes is smaller than $\beta$ or at high enough fields, the field $E_{[111]}$ and polarization $P_{[111]}$ along [111] are also expected to be related by equation~(\ref{cubic:OK}) but with modified coefficients.   As was shown by in \cite{Pertsev} and \cite{Catalan} the effect of substrate strain is also to  rescale the LGD coefficients. From equation~(\ref{cubic:OK})  the inverse susceptibility $\chi$, and the spontaneous polarization are given by
\begin{eqnarray}
P_s^2=-\frac{\alpha}{\beta}=\frac{(T_C-T)}{\beta\epsilon_0 C}, \;\;\;\;\label{Ps2:OK}\\ 
\frac{1}{\epsilon\epsilon_0}\approx \frac{1}{\chi}=-2\alpha=\frac{2(T_C-T)}{\epsilon_0 C}.
\label{InvEps:OK}
\end{eqnarray}
The plot of $P_r^2$ vs. $T$ is shown in figure~\ref{P_V:OK}(b) and was used to extract the values of $\beta '=1.52\times 10^9$ m$^5$C$^{-2}$F$^{-1}$ and $T_C '=620$~K; the latter  together with the previously obtained Curie-Weiss constant $C$ give the parameter $\alpha '$ as a function of temperature;  the primes are used to show that these are no longer the bulk LGD coefficients, but their effective values modified by strain, space charge and the fact that the samples are not [100] oriented. 

In another paper \cite{Zubko} we have studied phenomenologically the effects of space charge on hysteresis loops for fully and partially depleted samples with various dopant concentrations. Our results showed that increasing the amount of space charge caused a decrease in polarization and coercive field values as well as an increase in the observable dielectric constant. We also found that in the case of fully depleted films the effect of space charge was to simply rescale the original $\alpha$ and $\beta$ coefficients; the \emph{P-E} loops and the zero-field dielectric constant are completely depermined by these rescaled coefficients via (\ref{cubic:OK}) and (\ref{InvEps:OK}). This was not the case for partially depleted films, where the hysteresis loops  no longer fitted a cubic relationship at all.

The LGD fit using the experimentally extracted $\alpha '$ and $\beta '$ coefficients is shown in  figure~\ref{P_V:OK}(a) and is almost perfect, which would be consistent with the idea of a fully depleted sample, as would be expected for an insulating film.

\section{Discussion}
We have presented a large amount of data analyzed in various ways to try and find some clues as to the nature of the leakage mechanism in these commercial PZT capacitors. We now try to summarise  our findings and make some conclusions. 

SCL conduction has been ruled out as the \emph{I-V} do not seem to satisfy the required power law or the possible temperature dependences. We also rule out Fowler-Nordheim tunnelling precisely because of the strong temperature dependence; the \emph{I-V} also did not fit the $\log I/V^2\propto 1/V$ law for any significant voltage range.

Schottky thermionic emission assuming a uniform internal field distribution yields an unphysical value for the dielectric constant, as does the modified Schottky model proposed by Simmons \cite{Simmons} and used by Zafar \emph{et al.} \cite{Zafar, JimsBook} to fit their data on (Ba,Sr)TiO$_3$ thin films. Thus both these mechanisms can be ruled out. Fow completeness, we should note that the ``standard'' Schottky equation (\ref{Schottky:OK}) and that due to Simmons are just limiting cases of the more general thermionic emission-diffusion model by Crowell and Sze (see \cite{Sze, Baniecki}) which predicts the same \emph{J-E} expression as (\ref{Schottky:OK}) but with $A^*$ replaced by $A^{**}(E)$. Due to the unknown field dependence of $A^{**}$ it is difficult to fit data to this generalized model, however, this field dependence is usually much weaker than exponential and thus any functional deviations from (\ref{Schottky:OK}) are significant mainly at low fields. Our data spans a large range of biases and we have already ruled out the two limiting cases of the thermionic emission-diffusion model, thus it is probably safe to rule it out completely for the case of uniform fields on the basis of the unphysical value it gives for $\epsilon$ upon fitting.

If the sample is assumed to be partially depleted with the field at the potential maximum given by (\ref{FieldPartDepl:OK}) it is possible to obtain a fairly good fit to the Schottky equation (\ref{Schottky:OK}) with physically plausible parameters, at least over a limited range of applied biases and temperatures. However, the fits yield a fairly high  value for the concentration of ionised dopants that, according to our theoretical calculations \cite{Zubko}, is expected to produce a significant deformation of the hysteresis loop, which is not observed, and is also in fact expected to modify the internal field profile from a linear to a cubic, making equation~(\ref{FieldPartDepl:OK}) invalid and resulting in the loss of self-consistency in the analysis. Even without resorting to theoretical modelling, there seems to be a large discrepancy between the ionised dopant concentration of $10^{20}$--$10^{22}$cm$^{-3}$ extracted from \emph{I-V} data and the carrier concentration of $10^{18}$cm$^{-3}$ obtained from \emph{C-V} measurements. The mobile carrier density which can follow the ac signal used in \emph{C-V} measurements \emph{is} expected to be lower than $N_D$ \cite{Pintilie} but not by three or four orders of magnitude. The partial depletion model, although unlikely in this case, is very difficult to rule out definitively as it contains many adjustable parameters such as non-uniform dopant concentrations, unknown $m^*$, $V_{bi}$ and the field dependence of $A^{**}$ which cannot be independently measured.

The last possibility we have considered was Poole-Frenkel conduction which so far seems to fit all the experimental evidence available. The \emph{I-V} give very convincing straight lines on the $\log (I/V)$ vs $V^{1/2}$ plot for all temperatures and almost all two decades of applied bias. The extracted $\epsilon_\infty\approx 6.3$ shows good agreement with other independent measurements of the refractive index of PZT; and the assumption of a uniform field distribution is consistent with the shape of the \emph{P-E} loops according to phenomenological modelling. The \emph{I-V} also show little dependence on the polarity of the applied bias, again confirming that the leakage currents are probably bulk limited.

Poole-Frenkel conduction in  PZT  has been previously reported by Chen \emph{et al.} \cite{Chen_PF}. They observed highly symmetric \emph{I-V} despite having different top and bottom electrodes (Au and Pt respectively) with Poole-Frenkel conduction identified above 40kV/cm by the extracted dielectric constant value of 6.3, which is in perfect agreement with ours. Nagaraj \emph{et al.} \cite{Nagaraj} also observed this mechanism for 120--360nm thin films with epitaxial (La,Sr)CoO$_3$ electrodes. In their case, it was identified at fields above $\sim 100$kV/cm and yielded an activation energy of 0.5--0.6eV attributed by the authors to ionisation of Ti$^{4+}$ to  Ti$^{3+}$. 

Nagaraj \emph{et al.} refer the 0.5eV activation energy to a private communication. Previous papers by the referenced author \cite{Robertson}, however, state that the Ti$^{3+}$ centre lies at least 1eV below the conduction band edge; the value of 1eV for the electronic trap depth was reconfirmed in further publications \cite{Smyth, Smyth2}. The  ionisation energies of Pb ions \cite{Robertson} and oxygen vacancies \cite{Smyth2} are believed to be much lower compared with our value of $E_t\approx 1eV$, so electron trapping at Ti$^{4+}$ sites may after all be responsible for the observed conductivity, as suggested in \cite{Nagaraj}. $E_t$ of 1eV is also similar to the activation energy for ionic transport of oxygen vacancies \cite{Smyth}, however, the Poole-Frenkel model explicitly assumes a Coulombic attraction between a singly charged mobile  and an oppositely charged stationary species; doubly charged carriers such as oxygen vacancies would change the slope of the  $\log (I/V)$ vs $V^{1/2}$ plot. It is thus believed that we are observing electron or hole but not ionic conduction in these integrated PZT capacitors.

Before concluding we would like to mention two more points. The first relates to distinguishing Poole-Frenkel and Schottky conduction on the basis of the extracted dielectric constant. If the value extracted from the Poole-Frenkel plot agrees with the optical dielectric constant, while that from the Schottky plot does not, then it is likely that the mechamism is indeed Poole-Frenkel conduction. However, the reverse is not necessarily true. As was discussed by Simmons \cite{Simmons_PF} and later by Mark and Hartman \cite{MarkHartman}, in some cases a factor of $\frac{1}{2}$ can enter the exponential term containing $\beta_{PF}$ in equation~(\ref{PooleFrenkel:OK}) giving equal slopes for both Schottky and Poole-Frenkel plots.

Secondly, we note that our phenomenological arguments regarding the fitting of the \emph{P-V} loop with coefficients obtained from measurements of $\epsilon$ and $P_r$ are only valid if the field is applied along the polar axis. If the field is along [111] (as in our case) and the polarization is along the c-axis, then the main contribution to the measured $\epsilon$  will be from the larger a-axis dielectric constant $\epsilon_a$, whereas the measured polarization will be the projection of $P_3$ onto the [111] direction. In this case the extracted Landau coefficients should not be able to reproduce the \emph{P-V} loop. Thus is we assume that the fit in figure~\ref{P_V:OK}(a) is not just fortuitous, then our polarization must be pointing along [111]. This may be possible, despite the tetragonal composition, through strain effects or through field induced change of symmetry to rhombohedral as described by Bell \cite{Bell}. However, so far we have not directly measured the direction of the polarization in our samples, nor have we explored in much detail the effects of different field orientations in our phenomenological modelling. Here we merely wish to point out that \emph{P-V} data may be of use in providing clues as to the extent of depletion in ferroelectric thin films.

\section{Conclusions}
The leakage mechanism in commercial PZT capacitors has been studied in detail. It was shown that standard automated \emph{I-V} tests are not always appropriate for ferroelectrics due to the large difference in relaxation times for current transients at different fields. We have thus used current-time measurements to ensure that reliable steady state \emph{I-V}  were obtained. Together with \emph{C-V} these data were carefully analysed  and it was concluded that Poole-Frenkel emission of carriers from traps was the most likely conduction mechanism. A phenomenological model was used to confirm that the assumed field distribution was consistent with the observed shape of the ferroelectric hysteresis loop.

\section{Acknowledgements}
The authors are very grateful to Lucian Pintilie and Marin Alexe for their help with  the low temperature data, and to Finlay Morrison  and to Gustau Catalan for interesting discussions and many useful suggestions.

\end{document}